\documentclass[%
 reprint,
superscriptaddress,
 amsmath,amssymb,
 aps,
]{revtex4-2}

\usepackage{graphicx}
\usepackage{dcolumn}
\usepackage{bm}

\usepackage{physics}
\usepackage{subfigure}
\usepackage{lipsum}
\usepackage{CJK}
\begin{document}
\begin{CJK*}{UTF8}{gbsn}
\preprint{APS/123-QED}

\title{Measurement-device-independent quantum key distribution for non-standalone networks}
\author{Guan-Jie Fan-Yuan ({\CJKfamily{gbsn}范元冠杰})}
\author{Feng-Yu Lu ({\CJKfamily{gbsn}卢奉宇})}
\author{Shuang Wang ({\CJKfamily{gbsn}王双})}
\email{wshuang@ustc.edu.cn}
\author{Zhen-Qiang Yin ({\CJKfamily{gbsn}银振强})}
\author{De-Yong He ({\CJKfamily{gbsn}何德勇})}
\author{Zheng Zhou ({\CJKfamily{gbsn}周政})}
\author{Jun Teng ({\CJKfamily{gbsn}滕俊})}
\author{Wei Chen ({\CJKfamily{gbsn}陈巍})}
\author{Guang-Can Guo ({\CJKfamily{gbsn}郭光灿})}
\author{Zheng-Fu Han ({\CJKfamily{gbsn}韩正甫})}

\affiliation{CAS Key Laboratory of Quantum Information, University of Science and Technology of China, Hefei, Anhui 230026, China}
\affiliation{CAS Center for Excellence in Quantum Information and Quantum Physics, University of Science and Technology of China, Hefei, Anhui 230026, China}
\affiliation{State Key Laboratory of Cryptology, P. O. Box 5159, Beijing 100878, China}

\date{\today}

\begin{abstract}
Untrusted node networks initially implemented by measurement-device-independent quantum key distribution (MDI-QKD) protocol is a crucial step on the roadmap of Quantum Internet. Considering extensive QKD implementations of trusted node networks, a workable upgrading tactic of existing networks toward MDI networks needs to be explicit. Here, referring to the non-standalone (NSA) network of 5G, we propose an NSA-MDI scheme as an evolutionary selection for existing phase-encoding BB84 networks. Our solution can upgrade the BB84 networks and terminals that employ various phase-encoding schemes to immediately support MDI without hardware changes. This cost-effective upgrade effectively promotes the deployment of MDI networks as a step of untrusted node networks while taking full advantage of existing networks. Besides, the diversified demands on security and bandwidth are satisfied and the network survivability is improved.
\end{abstract}

\maketitle
\end{CJK*}

\section{Introduction}
Quantum key distribution (QKD)\cite{bennet1984quantum, ekert1991quantum, pirandola2019advances} can share a private key securely between two authorized parties, Alice and Bob. This private key can establish unconditional secure communication combined with the one-time pad\cite{vernam1926cipher}. The security of QKD relies on the principles of quantum physics, with any eavesdropping on a quantum channel being detected inevitably by extra signal disturbance\cite{gisin2002quantum, gottesman2004security, scarani2009security, pirandola2017fundamental}. Comparing with classical cryptography, the security of QKD is independent of the computation complexity. Therefore, QKD is counted among the solutions to secure communication in the quantum age.

For large-scale applications, networking is imperative for QKD, which can provide secure commutation service for numerous users\cite{elliott2002building, frohlich2013quantum}. In recent years, many influential networks\cite{elliott2005current, peev2009secoqc, chen2009field, wang2010field, chen2010metropolitan, stucki2011long, sasaki2011field, wang2014field, chen2021an} are conducted, including mature demonstrations for real-life applications. These works mark the achievement of the trusted node network and take the initial step of Quantum Internet\cite{wehner2018quantum}.

The trusted node network cannot provide end-to-end QKD services without the credibility of intermediary nodes. This limitation lowers the survivability of networks which means the ability of the network to provide secure key-distribution service if there are trusted nodes are controlled by eavesdroppers. In trusted-node-based networks, if the loss of credibility happens to one node due to attacks, large parts of the network may be paralyzed (full connection is a solution but with high costs and low feasibility, Ref. \cite{zhou2019security} provides a practical and detailed analysis for the scenario). For example, an untrusted central node can deprive the star topology of function, and losing credibility of a relay node can split a line-topology network, where the star topology and the link topology are normally used to constructing quantum network\cite{frohlich2013quantum, chen2021an} in metropolitan and wide areas, respectively. Therefore, upgrading such networks to untrusted-node-based networks for moving away from dependence on node credibility is necessary.

Any schemes that support untrusted node can be theoretically employed in untrusted-node-based networks, such as quantum repeaters\cite{briegel1998quantum, duan2001long, sangouard2011quantum, bauml2015limitations}, measurement-device-independent QKD (MDI-QKD)\cite{braunstein2012side, lo2012measurement, curty2014finite}, twin-field QKD (TF-QKD)\cite{lucamarini2018overcoming, ma2018phase, wang2018twin, cui2019twin} and device-independent QKD (DI-QKD)\cite{mayers1998quantum, masanes2011secure, reichardt2013classical, vazirani2014fully, arnon2018practical}. Comparing with other schemes, MDI-QKD, the scheme that can completely remove all detector side-channel attacks and its measurement unit can be regarded as an untrusted node, is the most mature and easiest one to implement\cite{comandar2016quantum, wang2015phase, yin2016measurement, zhou2020experimental} and initially shows the capability of networking\cite{tang2016measurement}. Therefore, MDI-QKD is an immediate object of network upgrade\cite{wehner2018quantum}.

Although upgrading to MDI networks can improve survivability, the cost and demand must be emphasized in network upgrades. On the cost side, the main protocol in existing networks is the BB84 protocol, however, it is incompatible with the MDI protocol. One major difference is the measurement mechanism. MDI-QKD requires a Bell state measurement (BSM) in the measurement unit, but BB84 does not need that\cite{lo2014secure}. Therefore, the measurement unit of the BB84 protocol cannot be used in the MDI protocol. The other difference is in the encoder of the transmitter. The bases of state preparation in QKD are X, Y and Z which are corresponding to Pauli matrices $\sigma_x$, $\sigma_y$ and $\sigma_z$, respectively. Consider the example of phase encoding in fibre-based QKD. In the BB84 protocol, all three bases can reach a low error rate, any two of them can be used. However, in MDI protocol, only Z basis (time-bin basis) can reach a low error rate, two employed bases must contain Z basis and the other basis is X or Y. Although the difference in encoder can be harmonized by constraining the basis choice of BB84 protocol or post-selection technique\cite{ma2012alternative}, the minimum requirement of the upgrade is to replace all QKD receivers, which is still a significant expenditure for device manufacturers and users. On the demand side, the requirements of two types of communication channels, control channel and data channel, in security and bandwidth are different. The control channel which transfers command messages between devices requires high security but low bandwidth. By contrast, the data channel requires high bandwidth but a relative low-security level. These two channel scenarios are suitable for MDI-QKD and BB84 protocols, respectively, because the former is more secure than the latter but with a lower secure key rate\cite{yuan201810}. Therefore, considering the cost and demand, the hasty upgrade of existing networks is inadvisable, and a workable roadmap toward MDI networks needs to be explicit.

A viable solution to the cost and demand issues is the non-standalone (NSA) network in 5G\cite{nikolich2017standards}. In the deployment of 5G, the devices of 4G also faces the problem of being replaced. In addition, that few user devices that support 5G mode requires a balance between the progress of the deployment and the demand. The NSA architecture is a step-by-step transition from 4G to 5G. By changing parts of devices, the 5G technology can be initially supported in 4G networks. As a result, the 5G network will be established completely as all old devices are replaced. Contrasted with the standalone (SA) network, the NSA upgrades 4G networks while taking full advantage of them, which is the most economical evolution path. Therefore, the NSA can be transplanted to QKD networks.

The central feature of the NSA is supporting a new protocol on almost-old facilities. Regardless of the NSA, the presses have reported that supporting MDI on polarization-encoding SARG04\cite{mizutani2014measurement} and BB84\cite{qi2015free} and demonstrating the polarization-encoding reconfigurable network of QKD and quantum digital signature (QDS)\cite{roberts2017experimental}. However, the solution for phase-encoding systems is still missing, which is probably more relevant because the phase encoding has an advantage of tolerance to channel disturbance which helps it to be widely deployed in established QKD networks\cite{elliott2005current, peev2009secoqc, chen2009field, wang2010field, stucki2011long, sasaki2011field, wang2014field} and become a mature commercialized solution. Therefore, the issues of the upgrade are imperative to phase-encoding networks, and the NSA can be harnessed in the upgrade toward MDI networks.

Here we propose an NSA-MDI scheme as the evolution of existing phase-encoding BB84 networks which harmonizes BB84 and MDI protocols in a single system. In our design, the MDI protocol can be implemented based on existing phase-encoding BB84 networks with few hardware changes. The barriers to the deployment of MDI networks is lowered, and the maximum utilization of existing networks shows the cost-effective side of this design. In addition, the supportability of MDI and BB84 protocols not only enables the utilization of the high key rate of BB84 and the high practical security of MDI in a single network for various application scenarios but protects existing trusted-node-based BB84 networks from paralysis when the credibility of some nodes is lost. Such advantages over cost and demand benefit the manufacturers, the service providers and the users and then promote the deployment of MDI networks.

To achieve that, the incompatibilities of encoding and measurement between MDI and BB84 protocols must be removed. Here we redesign the optical structure of the BSM unit in MDI protocol by introducing the same structure as original BB84 systems, which is not limited to a specific realization of phase encoding. In this paper, an asymmetric Mach-Zehnder (MZ) structure\cite{gobby2004quantum, yuan2007unconditionally, zhang2018proof} and an asymmetric Faraday-Michelson (FM) structure\cite{mo2005faraday, liu2019practical, wang2014field} are adopted as an example shown in Sec. Protocol and Sec. Experimental System, respectively. The new BSM of the MDI protocol is identical to the decoder of the BB84 protocol. Only a polarization controller is added to realize polarization indistinguishability for MDI protocol if it does not exist in the BB84 system using, for example, FM structures. Therefore, our evolution path can offer existing BB84 networks the capacity of MDI protocol with few hardware cost. In addition, the theoretical error rate of X basis can be low in our scheme (detailed theoretical proof is shown in Appendix \ref{app:scheme}), which avoids the need for optical switch\cite{tamaki2012phase} or phase-post-selection technique\cite{ma2012alternative}. To show the superiority of our evolution path, we experimentally demonstrate the BB84 protocol between Alice and Charlie, Bob and Charlie, and the MDI protocol between Alice and Bob, respectively. The experimental results are shown in Sec. Experimental Results. A summary is provided in Sec. Discussion.

\section{Protocol}
\label{sec:prot}
\begin{figure}[htbp]
\centering
\includegraphics[width=1\linewidth]{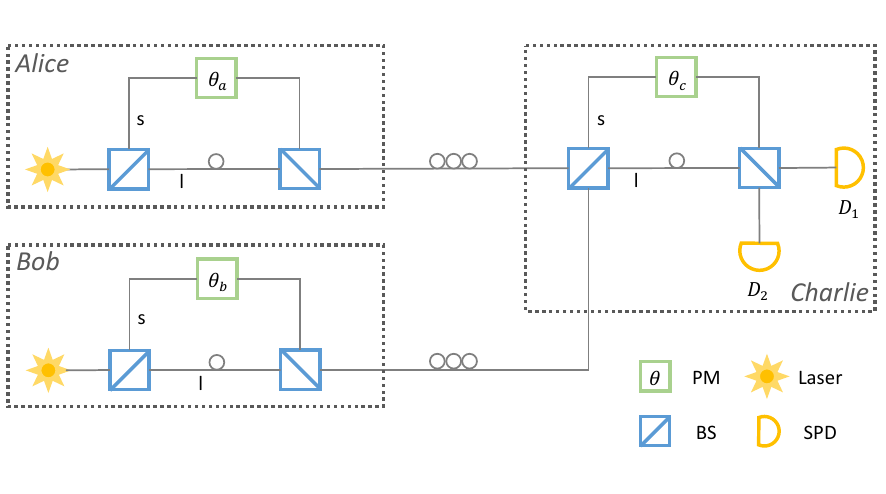}
\caption{Schematic diagram of the non-standalone MDI protocol. PM, phase modulator; Laser, pulsed weak-coherent source; BS, beam splitter; SPD, single photon detector.}
\label{fig:prot}
\end{figure}

Our MDI-QKD system is based on phase encoding and schematically shown in Fig. \ref{fig:prot}. Structurally, an asymmetric Mach-Zehnder interferometer (AMZI) is placed on Charlie to harmonize the MDI protocol with the BB84 protocol. This additional structure helps him not only communicate with Alice and Bob as a legal user using BB84 protocol, respectively, but also play the untrusted relay of MDI-QKD between Alice and Bob. It also can reduce the theoretical error rate of X basis in phase-encoding decoy-state MDI-QKD, which avoids extra payments of fast optical switch and phase-post-selection technique that the previous phase-encoding MDI-QKD schemes are needed to realize a phase shift compensation or a conversion from phase information to polarization information in Ref. \cite{tamaki2012phase} and reduce the error rate in the decoy-state protocols in Ref. \cite{ma2012alternative}, respectively.

In our system, the basis is chosen from $\mathcal{B}=\{X, Y\}$. X basis consisting of $\ket{+}=\frac{1}{\sqrt{2}}(\ket{s}+\ket{l})$ and $\ket{-}=\frac{1}{\sqrt{2}}(\ket{s}-\ket{l})$, and the Y basis consisting of $\ket{+i}=\frac{1}{\sqrt{2}}(\ket{s}+i\ket{l})$ and $\ket{-i}=\frac{1}{\sqrt{2}}(\ket{s}-i\ket{l})$, where $\ket{s}$ and $\ket{l}$ represent the time-bin states traveling along the short and long arm of AMZI, respectively. In addition, the three-intensity-decoy-state schemes\cite{wang2013three, wang2017measurement, lim2014concise} are adopted in both BB84 and MDI protocols. Specifically, The intensity of each laser pulse is randomly chosen from $\mathcal{I}=\{\mu, \nu, \omega\}$ and the intensities satisfy $\mu>\nu+\omega$ and $\nu>\omega\geq0$. We use $P^\beta_\iota$ to denote the probability that a laser pulse is prepared at a basis of $\beta\in\mathcal{B}$ and an intensity of $\iota\in\mathcal{I}$, respectively. For each instance of protocols, the intensities, $\mathcal{I}$ and the probabilities, $P^\beta_\iota$, are optimized for the maximum secure key rate.

\begin{figure*}[ht]
\centering
\includegraphics[width=1\textwidth]{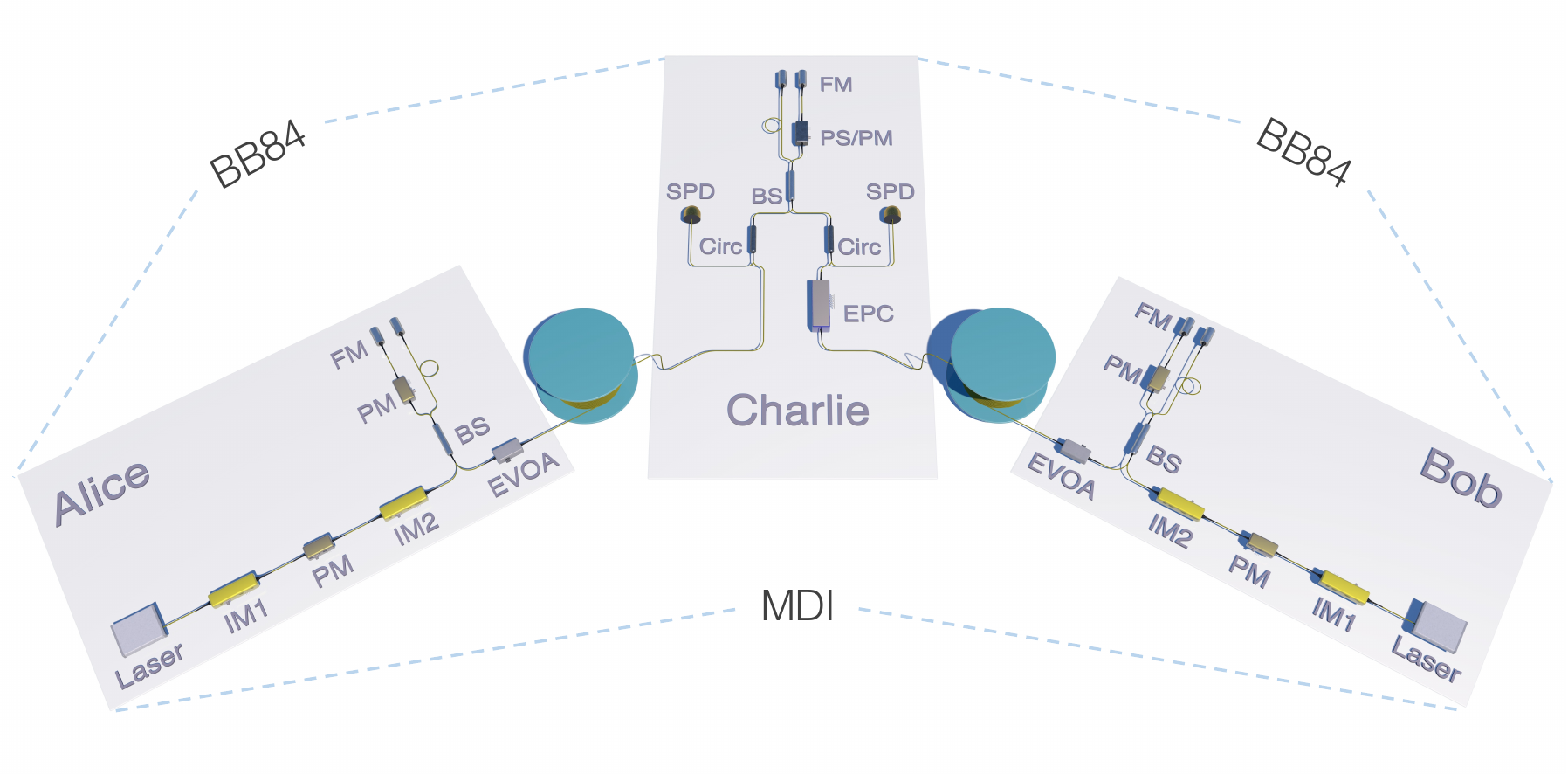}
\caption{Experimental setup for the non-standalone MDI-QKD system. Alice and Bob can implement phase-encoding MDI-QKD and generate secure key with Charlie via BB84. Laser, frequency-locked lasers; IM1, intensity modulator as pulse generator; IM2, intensity modulator as decoy state generator; BS, beam splitter; PM, phase modulator; PS, phase shifter; FM, Faraday mirror; EVOA, electronic variable optical attenuator; EPC, electronic polarization controller; Circ, circulator; SPD, single photon detector. For Alice, Bob and Charlie, the combination of one beam splitter, one phase controller and two Faraday mirrors constitutes their own asymmetric Faraday-Michelson interferometer (AFMI), the other PM is used to phase randomization.}
\label{fig:sch}
\end{figure*}

\subsection{MDI}
For MDI protocol, Alice and Bob randomly prepare their quantum state on bases $\beta_a$ and $\beta_b$, respectively. Each inceptive pulse is divided into two adjacent pulses by their AMZIs. A relative phase is introduced by the phase modulator of AMZI according to the selected basis and key. The relations of the modulated phase to basis and key are listed in Table \ref{tab:bkmdi} where $\theta_a$ and $\theta_b$ represent the relative phase modulated by Alice and Bob, respectively. In addition, the intensities of their laser pulses are modulated as $\iota_a$ and $\iota_b$, respectively, and the corresponding probabilities are $P^{\beta_a}_{\iota_a}$ and $P^{\beta_a}_{\iota_a}$, respectively.

\begin{table}[h]
\centering
\caption{\bf The code table in MDI protocol}
\label{tab:bkmdi}
\setlength{\tabcolsep}{18pt}{
\begin{tabular}{ccccc}
\hline
 & $\ket{+}$ & $\ket{-}$ & $\ket{+i}$ & $\ket{-i}$ \\
\hline
$\theta_a$ & $0$ & $\pi$ & $\frac{\pi}{2}$ & $\frac{3\pi}{2}$ \\
$\theta_b$ & $0$ & $\pi$ & $\frac{\pi}{2}$ & $\frac{3\pi}{2}$ \\
\hline
\end{tabular}}
\end{table}

Then Alice and Bob send their pulses to Charlie for BSM. For each pair of quantum states, unlike the original, the AMZI of Charlie further divides the incident pulses into three states of timestamps, $\ket{ss}$, $\ket{sl(ls)}$ and $\ket{ll}$. Charlie detects the middle one with two single-photon detectors (SPDs) because it contains the phase information of Alice and Bob. After basis sifting, the coincidence counting is retained which represent a successful BSM, and other events are discarded. Here we briefly show the probabilities of valid responses, $Q$, on the conditions that $\theta_a=\theta_b$ and $\theta_a\neq\theta_b$ and the error rates, $E$.
\begin{equation}
\begin{aligned}
&\eval{Q^X_{\iota_a\iota_b}}_{\theta_a=\theta_b}=\frac{\iota_a\iota_b}{2} \\
&\eval{Q^X_{\iota_a\iota_b}}_{\theta_a\neq\theta_b}=0 \\
&E^X_{\iota_a\iota_b}=\frac{\eval{Q^X}_{\theta_a\neq\theta_b}}{\eval{Q^X}_{\theta_a=\theta_b}+\eval{Q^X}_{\theta_a\neq\theta_b}}=0 \\
&\eval{Q^Y_{\iota_a\iota_b}}_{\theta_a=\theta_b}=\frac{(\iota_a+\iota_b)^2-2\iota_a\iota_b}{8} \\
&\eval{Q^Y_{\iota_a\iota_b}}_{\theta_a\neq\theta_b}=\frac{(\iota_a+\iota_b)^2+2\iota_a\iota_b}{8} \\
&E^Y_{\iota_a\iota_b}=\frac{\eval{Q^Y}_{\theta_a=\theta_b}}{\eval{Q^Y}_{\theta_a=\theta_b}+\eval{Q^Y}_{\theta_a\neq\theta_b}}=\frac{\iota_a^2+\iota_b^2}{2(\iota_a+\iota_b)^2}\overset{\iota_a=\iota_b}{=}\frac{1}{4} \\
\label{qexy}
\end{aligned}
\end{equation}
In these equations, for simplification, the dark-count rate and afterpulse probability of detector are neglected, the detection efficiency and transmittance are regarded as 100\%, and the reference frames of Alice and Bob are aligned which means $\theta_c=0$. The realistic version is detailedly shown in Appendix \ref{app:scheme}. Note that because $\eval{Q^X}_{\theta_a\neq\theta_b}=0$, we regard the responses under $\theta_a=\theta_b$ and $\theta_a\neq\theta_b$ as correct and error responses of X basis, respectively. On the contrary, for Y basis, the responses under $\theta_a\neq\theta_b$ is regarded as correct responses for a low error rate.

According to the Eqs. (\ref{qexy}), the error rate of the X basis can be very low which reflects the same characteristic with the Z basis used in the original scheme\cite{wang2015phase, yin2016measurement, wang2017measurement}. Therefore, our scheme can realize a phase-encoding MDI protocol by only modulating the phase. Neither optical switch nor phase-post-selection technique is not required. Such an encoding scheme is also consistent with the phase-encoding BB84 protocol.

Finally, with the data of $Q^{\beta}_{\iota_a\iota_b}$ and $E^{\beta}_{\iota_a\iota_b}$, the secure key can be extracted from the data when both Alice and Bob encode their bits using signal states ($\mu$) on the X basis. The single-photon yield and error yield can be estimated by the rest of the data and engaged in the calculation of the secure key rate\cite{wang2017measurement}. The SKR is given by
\begin{equation}
R={P_\mu^X}^2(\mu^2e^{-2\mu}Y_{11}^{X,L}(1 - H_2(e_{11}^{Y,U}))-Q^X_{\mu_a \mu_b} f_eH_2((E^X_{\mu_a \mu_b}))
\end{equation}
where $Y_{11}^{Z,L}$ is the lower bound of the yield of single-photon pairs, $e_{11,p}^{Z,U}$ is the upper bound of the phase-flip error rate, $H_2(x)=-x log_2(x)-(1-x)log_2(1-x)$ is the binary Shannon entropy function and $f_e$ is the error correction efficiency. The calculation of secure key rate are also detailedly shown in Appendix \ref{app:mdirate}.

\subsection{BB84}
For BB84 protocol, both Alice and Bob can communicate with Charlie. Here we provide a detailed description of the protocol by the example of the communication between Alice and Charlie. Alice first prepares her quantum state at a basis $\beta_a$ and an intensity $\iota_a$ with probabilities $P^{\beta_a}_{\iota_a}$. The AMZI also divides each laser pulse into two adjacent pulses. The phase modulators modulate the relative phase, $\theta_a$, between them according to the random basis and key Alice selected. The relations of the modulated phase to basis and key are listed in Table \ref{tab:bkbb84}.
\begin{table}[h]
\centering
\caption{\bf The code table in in BB84 protocol}
\label{tab:bkbb84}
\setlength{\tabcolsep}{16pt}{
\begin{tabular}{ccccc}
\hline
 & $\ket{+}$ & $\ket{-}$ & $\ket{+i}$ & $\ket{-i}$ \\
\hline
$\theta_a(\theta_b)$ & $0$ & $\pi$ & $\frac{\pi}{2}$ & $\frac{3\pi}{2}$ \\
$\theta_c$ & $0$ & $0$ & $\frac{\pi}{2}$ & $\frac{\pi}{2}$ \\
\hline
\end{tabular}}
\end{table}

Then Alice sends her quantum state to Charlie. Charlie selects a basis $\beta_c$ with probabilities $P^\beta_c$ by modulating the relative phase, $\theta_c$. The relation between $\theta_c$ and basis is also listed in Table \ref{tab:bkbb84}. After the transmission along Alice's and Charlie's AMZIs, the original pulse is split into three parts according to different paths: two short arms ($\ket{ss}$), one short and one long arm ($\ket{sl(ls)}$), two long arms ($\ket{ll}$). Charlie detects the $\ket{sl(ls)}$ with two single-photon detectors (SPDs). In all possible outcomes, no detection events are discarded, the others are counted as valid responses if the bases of Alice and Charlie are identical. Specially, due to the random assignment of a bit value, the double-click events cause 50\% of error rates\cite{jain2011device}. Then, the yield, $Q^\beta_\iota$, and error rate, $E^\beta_\iota$, of a basis $\beta$ and an intensity $\iota$ can be obtained from statistics.

Finally, with the data of $Q^\beta_\iota$ and $E^\beta_\iota$, the parameters which are required in the calculation of the secure key rate can be estimated and bounded by decoy technology\cite{lim2014concise}. The secure key rate can be obtained by
\begin{equation}
R=\frac{1}{N}(s_0^X+s_1^X(1-H_2(e_{1,p}^X))-\lambda_{EC}-6\log_2\frac{21}{\varepsilon_{sec}}-\log_2\frac{2}{\varepsilon_{cor}})
\end{equation}
where $s_0$ is the number of vacuum events, $s_1$ is the number of single-photon events, $e_{1,p}$ is the phase error rate, $N$ is the total number of pulses (sent by Alice or Bob), $H_2(x)=-x log_2(x)-(1-x)log_2(1-x)$ is the binary Shannon entropy function, $\lambda_{EC}=n^\beta f_{e}H_2(E^\beta)$ is the consumption of the information in error-correction, $f_{e}$ is the efficiency factor of the error-correction method used, $\varepsilon_{cor}$ and $\varepsilon_{sec}$ are secure parameters. The detail of the calculation of secure key rate are shown in Appendix \ref{app:bb84rate}.

\section{Experimental System}
\label{sec:exp}
The experimental setup is schematically introduced as shown in Fig. \ref{fig:sch}. The laser and intensity modulator (IM) 1 of Alice and Bob compose the weak-coherent pulse source. The laser is a frequency-locked continuous-wave source whose central wavelength is locked to a molecular absorption line at 1542.38 nm, with a precision of 0.0001 nm, corresponding to an approximately 10 MHz accuracy in the spectrum domain. Then, IM1 chops the continuous-wave laser into pulses with a 2.5 ns temporal width and 40 MHz repetition rate.

IM2 and an electronic variable optical attenuator (EVOA) of Alice and Bob modulate the intensity of pulses for decoy-state technology\cite{lo2005decoy, wang2005beating, ma2005practical, zhou2016making, fan2021optimizing}. In our system, the three-intensity decoy-state method is used and these three intensities are denoted by $\mu, \nu, \omega$, respectively. IM2 modulates the intensity according to the decoy-state method. The EVOA enables single-photon attenuation of the modulated pulses.

The asymmetric Faraday-Michelson interferometers (AFMIs) replace the AMZIs for robustness and are used for the phase-encoding quantum state preparation. The two arms of an AFMI are called the short arm ($s$) and the long arm ($l$), respectively. Specifically, for each AFMI, each laser pulse is split into two adjacent pulses by the beam splitter (BS). Then the phase modulator (PM) modulates the relative phase between them. Here the modulated relative phase of Alice, Bob and Charlie are denoted by $\theta_a$, $\theta_b$ and $\theta_c$, respectively, and $\theta\in\{0, \pi, \pi/2, 3\pi/2\}$ which correspond to $\ket{+}, \ket{-}, \ket{+i}, \ket{-i}$, respectively. The Faraday mirror (FM) rotates the polarization to compensate the disturbances caused by birefringence within the asymmetric Faraday-Michelson interferometer.

Charlie is linked to Alice and Bob by a 10-kilometre length of optical fibre, respectively, corresponding to 1.96 dB of loss for each link. The electronic polarization controller (EPC) of Charlie is used to guarantee the polarization indistinguishability of Alice's and Bob's states in the MDI protocol. A dual-EPC configuration\cite{tang2016measurement} or polarization scrambling method\cite{wang2017measurement} is more efficient in field environment, which is a tradeoff between performance and cost. After the decoding of Charlie's FMI, the laser pulses become a superposition of three timestamps state corresponding to the paths of laser pulses. In both BB84 and MDI protocols, the middle pulses are detected for Mach-Zehnder interference and BSM, respectively, by two InGaAs/InP single-photon detectors (Qasky WT-SPD300-LN\cite{qasky2018}) with a detection efficiency of 25\% and an averaged dark count rate of $7.5\times10^{-6}$ per gate. Moreover, the internal transmittance of Charlie's optical components is 4.2 dB.

In electronics, all intensity modulators and phase modulators are driven by homemade digital-to-analogue converters (DACs). The modulating voltages of them are found by scanning the outputs of DACs. Specifically, for the three AFMIs, the voltages of the modulated relative phases can be obtained by alternately modulating two different code of the phase modulator while maintaining the voltages of other modulators. Scanning the voltage difference between the two coding modes and recording the counts' curves of one detector in the two coding modes, the phase difference between the curves is the differential phase to which the differential voltage corresponds. Specially, when conducting the MDI protocol, the phase modulator of Charlie can be used as a phase shifter to compensate reference-frame misalignment between Alice and Bob by scanning the voltage to minimize the coincidence counts when they select different phase in X basis or the same phase in Y basis. The system is controlled by an FPGA module (NI PCIe-7852R). The FPGA module converts the random information of basis and key to digital signals and then sends them to the DACs for encoding. The FPGA module also implements basis sifting and data collection according to the encoding information and responses of SPDs.

\section{Experimental Results}
\label{sec:res}
\subsection{MDI QKD}
\label{sec:mdi}
We first test the system's performance by measuring the visibility of Hong-Ou-Mandel (HOM) interference on Charlie's side. We obtain a visibility of 47.8\% over 20 km of single-mode fiber which approaches to theoretical limit of 50\% for weak coherence sources.

For high performance, we optimize the parameters of our system before key distribution using Particle Swarm Optimization (PSO)\cite{kennedy2010particle}. Specifically, $\mu=0.284, \nu=0.057, \omega=0, P_\mu^X=0.466, P_\mu^Y=0.035, P_\nu^X=0.076, P_\nu^Y=0.293, P_\omega=0.130$.

Then, to achieve a higher practical security, we consider finite-size effect and adopt three-intensity decoy-state method. Here we apply the large deviation theory, specifically, the Chernoff bound\cite{chernoff1952measure, curty2014finite}, for the fluctuation estimation in our experiment, with a fixed failure probability of $\varepsilon=10^{-10}$ and a total number of sifted pulse pairs $N_t=3.97\times10^{11}$. Finally, we obtain the secure key rate of $1.025\times10^{-5}$ for transmission distance of 20 km as shown in Fig. \ref{fig:res}. Then the gains and quantum bit error rates (QBERs) of our MDI-QKD system are shown in Table \ref{tab:mdires}. The method to calculate the secure key rate is shown in Appendix \ref{app:mdirate}.
\begin{table}[h]
\centering
\caption{\bf The experimental gains and quantum bit error rates of our MDI-QKD system}
\label{tab:mdires}
\setlength{\tabcolsep}{8pt}{
\begin{tabular}{ccccc}
\hline
$\mu_a\mu_b$ &$Q^X$ & $E^X$ & $Q^Y$ & $E^Y$ \\
\hline
$\mu\mu$ & $1.82\times10^{-4}$ & $2.69\%$ & $3.40\times10^{-4}$ & $26.08\%$ \\
$\mu\nu$ & $4.67\times10^{-5}$ & $4.75\%$ & $1.19\times10^{-4}$ & $36.26\%$ \\
$\mu\omega$ & $7.06\times10^{-6}$ & $51.02\%$ & $1.06\times10^{-4}$ & $50.10\%$ \\
$\nu\mu$ & $4.90\times10^{-5}$ & $5.02\%$ & $1.32\times10^{-4}$ & $35.86\%$ \\
$\nu\nu$ & $1.11\times10^{-5}$ & $3.66\%$ & $2.13\times10^{-5}$ & $26.16\%$ \\
$\nu\omega$ & $2.39\times10^{-7}$ & $47.26\%$ & $5.63\times10^{-6}$ & $50.46\%$ \\
$\omega\mu$ & $4.04\times10^{-6}$ & $50.19\%$ & $1.01\times10^{-4}$ & $50.40\%$ \\
$\omega\nu$ & $9.25\times10^{-7}$ & $51.88\%$ & $4.80\times10^{-6}$ & $50.09\%$ \\
\hline
\end{tabular}}
\end{table}

\subsection{BB84 QKD}
\label{sec:bb84}
Similarly, we test the system's performance by measuring the visibility of MZ interferometer. We obtain the visibilities of 99.7\% (Alice-Charlie) and 99.5\% (Bob-Charlie) over 10 km of single-mode fiber which approaches to theoretical limit of 100\% for weak coherence sources.

Also, we first optimize the parameters of BB84 systems. For simplicity, the parameters of Alice-Charlie system are the same as Bob-Charlie system's. Specifically, $\mu=0.538, \nu=0.063, \omega=0.003, P_\mu^X=0.531, P_\mu^Y=0.110, P_\nu^X=209, P_\nu^Y=0.043, P_\omega^X=089, P_\omega^Y=0.018$.

Then, for a higher practical security, we implement the security analysis in \cite{lim2014concise} and three-intensity decoy-state method. The Hoeffding's inequality\cite{hoeffding1994probability} is used for our fluctuation analysis. In our calculation of secure key rate, the failure probability of parameter estimation, $\varepsilon_{sec}$, is equal to $10^{-9}$ and the failure probability of error-verification step, $\varepsilon_{cor}$, is equal to $10^{-15}$. Besides, the total number of sifted pulse pairs $N_t=1.16\times10^{9}$.

Finally, we obtain the secure key rates of $6.289\times10^{-3}$ (Alice-Charlie) and $6.155\times10^{-3}$ (Bob-Charlie) for transmission distance of 10 km as shown in Fig. \ref{fig:res}. Then the gains and QBERs of our BB84 QKD systems are shown in Table \ref{tab:bb84res}. The method to calculate the secure key rate is shown in Appendix \ref{app:bb84rate}.
\begin{table}[h]
\centering
\caption{\bf The experimental gains and quantum bit error rates of our BB84 QKD systems}
\label{tab:bb84res}
\setlength{\tabcolsep}{10pt}{
\begin{tabular}{ccccc}
\hline
$\mu_a$ &$Q^X$ & $E^X$ & $Q^Y$ & $E^Y$ \\
\hline
$\mu$ & $3.10\times10^{-2}$ & $0.39\%$ & $3.09\times10^{-2}$ & $0.28\%$ \\
$\nu$ & $3.67\times10^{-3}$ & $0.38\%$ & $3.71\times10^{-3}$ & $0.29\%$ \\
$\omega$ & $1.96\times10^{-4}$ & $2.66\%$ & $1.93\times10^{-4}$ & $1.82\%$ \\
\hline
$\mu_b$ &$Q^X$ & $E^X$ & $Q^Y$ & $E^Y$ \\
\hline
$\mu$ & $3.13\times10^{-2}$ & $0.38\%$ & $3.14\times10^{-2}$ & $0.34\%$ \\
$\nu$ & $3.69\times10^{-3}$ & $0.50\%$ & $3.71\times10^{-3}$ & $0.48\%$ \\
$\omega$ & $1.95\times10^{-4}$ & $1.89\%$ & $1.98\times10^{-4}$ & $2.09\%$ \\
\hline
\end{tabular}}
\end{table}

\subsection{Summary}
\begin{figure}[htbp]
\centering
\includegraphics[width=1\linewidth]{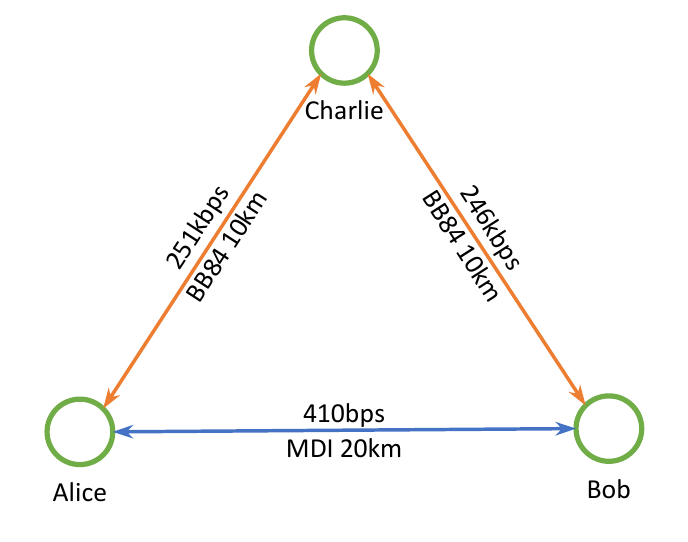}
\caption{Virtual network topology and link rates of our system.}
\label{fig:res}
\end{figure}

According to the results shown in Sec. MDI QKD and Sec. BB84 QKD We summarize the performance of our system as network link rates in Fig. \ref{fig:res}.

In summary, our system implements a compatibility between two distinct protocols: BB84 and MDI. The finite-size effect is included in our analysis for the requirement of practical security. Moreover, three-intensity decoy-state method is used. Based on the FM structure, our BB84 subsystems can automatically compensate for the channel polarization disturbance. The secure key rate is about 250 kbps at 10 km of fiber. Also, our new scheme can realize low-error-rate phase-encoding MDI-QKD without optical switch and phase-post-selection technique. The system complexities of Alice and Bob are reduced whose sensitivity to cost is higher than Charlie. And we obtain 410 bps of secure key rate with a high level of practical security.

\section{Discussion}
\label{sec:con}
In conclusion, with the help of our new scheme, the incompatibilities between phase-encoding MDI and BB84 protocols are removed. These protocols are integrated into a single system to introduce the advantages of the NSA network so that our system can switch the engaged protocol between them as required and further bridges the gap between existing phase-encoding BB84 networks and MDI networks.

BB84 is one of the most widely used protocols in trusted node networks, and MDI is an ideal candidate for the untrusted-node-based network. These features certainly show an evolution path toward MDI networks. The NSA-MDI scheme can immediately make the phase-encoding BB84 networks support MDI and is not limited to a specific realization of phase encoding. The dependence of networks on node credibility is also lowered, thus the network survivability is improved. Moreover, various requirements of different application scenarios, especially the requirements of high key rate or high-security level can be satisfied in one network. More importantly, all these advantages can be obtained without hardware cost which benefits from the coordination of our scheme.

The network is the final form of QKD application and the first step of Quantum Internet. During the popularization and upgrade of QKD, the cost needs to be lowered. Our NSA network scheme provides an evolution path, which targets both cost and demand. Existing phase-encoding BB84 networks and production lines are fully exploited, and the lower threshold and higher usability can quicken the construction of MDI-QKD networks.

\begin{acknowledgments}
This work was supported by the National Key Research And Development Program of China (Grant No. 2018YFA0306400), the National Natural Science Foundation of China (Grants No. 61622506, No. 61575183, No. 61627820, No. 61475148, and No. 61675189), the Project funded by China Postdoctoral Science Foundation (Grant No. 2021M693098), and the Anhui Initiative in Quantum Information Technologies.\end{acknowledgments}

\appendix

\section{MDI-QKD with phase-randomized coherent states}
\label{app:scheme}
\setcounter{equation}{0}
\renewcommand{\theequation}{A{\arabic{equation}}}
\setcounter{table}{0}
\renewcommand{\thetable}{A\arabic{table}}
\setcounter{figure}{0}
\renewcommand{\thefigure}{A\arabic{figure}}

\begin{figure}[htbp]
\centering
\includegraphics[width=1\linewidth]{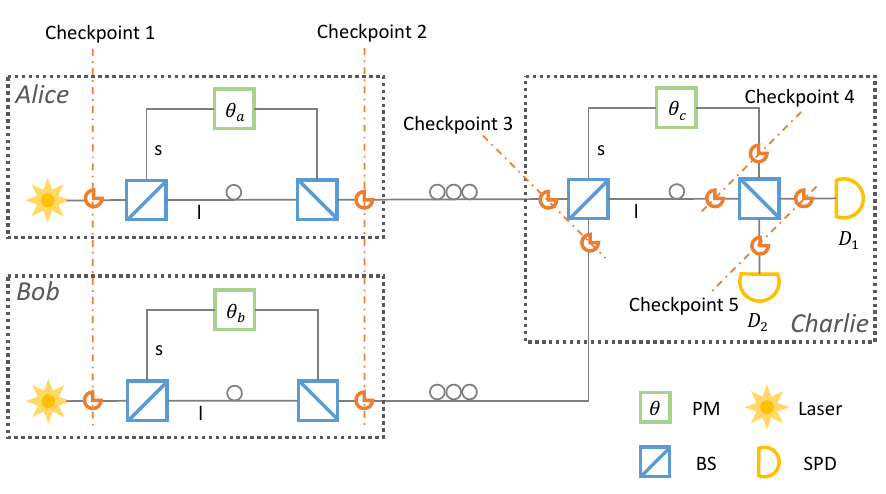}
\caption{Schematic diagram of the non-standalone MDI protocol with checkpoints.}
\label{fig:protcp}
\end{figure}

In this section, we detailedly shows the evolution of quantum state in our system according to the checkpoints marked on Fig. \ref{fig:protcp}.

\emph{Checkpoint 1.} Alice and Bob prepare coherent states with intensities $\mu_a$ and $\mu_b$, respectively, and randomize the phases. The initial joint state is 
\begin{equation}
\ket{e^{i\phi_a}\sqrt{\mu_a}}_a\ket{e^{i\phi_b}\sqrt{\mu_b}}_a
\end{equation}
where $\phi_a$ and $\phi_b$ are the overall randomized phases.

\emph{Checkpoint 2.} The pulses are split into two orthogonal optical modes (l mode and s mode). 
\begin{equation}
\ket{e^{i\phi_a}\sqrt{\frac{\mu_a}{2}}}_{a_l}\ket{e^{i(\phi_a+\theta_a)}\sqrt{\frac{\mu_a}{2}}}_{a_s}\ket{e^{i\phi_b}\sqrt{\frac{\mu_b}{2}}}_{b_l}\ket{e^{i(\phi_b+\theta_b)}\sqrt{\frac{\mu_b}{2}}}_{b_s}
\end{equation}
where $\theta_a$ and $\theta_b$ are the relative phases between the two modes which are modulate by Alice and Bob, respectively.

\emph{Checkpoint 3.} After passing through lossy channels, the joint state can be expressed by
\begin{equation}
\begin{aligned}
&\ket{e^{i\phi_a}\sqrt{\frac{\mu_a\eta_a}{2}}}_{a_l}\ket{e^{i(\phi_a+\theta_a)}\sqrt{\frac{\mu_a\eta_a}{2}}}_{a_s} \\
&\otimes\ket{e^{i\phi_b}\sqrt{\frac{\mu_b\eta_b}{2}}}_{b_l}\ket{e^{i(\phi_b+\theta_b)}\sqrt{\frac{\mu_b\eta_b}{2}}}_{b_s}
\end{aligned}
\end{equation}
where $\eta_a$ and $\eta_b$ are the channel transmittances of Alice-Charlie and Bob-Charlie.

\emph{Checkpoint 4.} The first BS and PM of Charlie transform the states into
\begin{equation}
\begin{aligned}
&\ket{e^{i\phi_a}\frac{\sqrt{\mu_a\eta_a}}{2}-e^{i\phi_b}\frac{\sqrt{\mu_b\eta_b}}{2}}_{ll} \\
&\otimes\ket{e^{i(\phi_a+\theta_a+\theta_c)}\frac{\sqrt{\mu_a\eta_a}}{2}+e^{i(\phi_b+\theta_b+\theta_c)}\frac{\sqrt{\mu_b\eta_b}}{2}}_{ss} \\
&\otimes\ket{e^{i(\phi_a+\theta_c)}\frac{\sqrt{\mu_a\eta_a}}{2}+e^{i(\phi_b+\theta_c)}\frac{\sqrt{\mu_b\eta_b}}{2}}_{ls} \\
&\otimes\ket{e^{i(\phi_a+\theta_a)}\frac{\sqrt{\mu_a\eta_a}}{2}-e^{i(\phi_b+\theta_b)}\frac{\sqrt{\mu_b\eta_b}}{2}}_{sl}
\end{aligned}
\end{equation}
where $\theta_c$ is the modulated phase of Charlie.

\emph{Checkpoint 5.} Before the pulses arrive at SPDs, the states are changed to
\begin{equation}
\begin{aligned}
&\ket{\frac{\sqrt{\mu_a\eta_a}}{2\sqrt{2}}(e^{i(\phi_a+\theta_c)}+e^{i(\phi_a+\theta_a)})+\frac{\sqrt{\mu_b\eta_b}}{2\sqrt{2}}(e^{i(\phi_b+\theta_c)}-e^{i(\phi_b+\theta_b)})}_{D_1} \\
&\otimes\ket{\frac{\sqrt{\mu_a\eta_a}}{2\sqrt{2}}(e^{i(\phi_a+\theta_c)}-e^{i(\phi_a+\theta_a)})+\frac{\sqrt{\mu_b\eta_b}}{2\sqrt{2}}(e^{i(\phi_b+\theta_c)}+e^{i(\phi_b+\theta_b)})}_{D_2}
\end{aligned}
\end{equation}

For simplicity, we use $\ket{\psi_1}_{D_1}$ and $\ket{\psi_2}_{D_2}$ to replace the expression above.
\begin{equation}
\begin{aligned}
&\ket{\psi_1}_{D_1}\otimes\ket{\psi_2}_{D_2}
\end{aligned}
\end{equation}

Then, the response probabilities of $D_1$ and $D_2$ can be obtained by
\begin{equation}
\begin{aligned}
&p_\mu^{D_1}=1-(1-Y_0)(1-P_{ap})\exp(-\abs{\psi_1}^2) \\
&p_\mu^{D_2}=1-(1-Y_0)(1-P_{ap})\exp(-\abs{\psi_2}^2) \label{pd}
\end{aligned}
\end{equation}
where $Y_0$ is the dark-count rate and $P_{ap}$ is the afterpulse rate of detectors\cite{fan2018afterpulse}.
\begin{equation}
\begin{aligned}
\abs{\psi_1}^2=&\frac{\mu_a\eta_a}{4}(1+\cos(\theta_c-\theta_a))+\frac{\mu_b\eta_b}{4}(1-\cos(\theta_c-\theta_b)) \\
&+\frac{\sqrt{\mu_a\eta_a\mu_b\eta_b}}{4}(\cos(\phi_a-\phi_b)+\cos(\phi_a-\phi_b+\theta_a-\theta_c) \\
&-\cos(\phi_b-\phi_a+\theta_b-\theta_c)-\cos(\phi_a-\phi_b+\theta_a-\theta_b)) \\
\abs{\psi_2}^2=&\frac{\mu_a\eta_a}{4}(1-\cos(\theta_c-\theta_a))+\frac{\mu_b\eta_b}{4}(1+\cos(\theta_c-\theta_b)) \\
&+\frac{\sqrt{\mu_a\eta_a\mu_b\eta_b}}{4}(\cos(\phi_a-\phi_b)+\cos(\phi_a-\phi_b+\theta_c-\theta_b) \\
&-\cos(\phi_b-\phi_a+\theta_c-\theta_a)-\cos(\phi_a-\phi_b+\theta_a-\theta_b)).
\end{aligned}
\end{equation}

For simplicity, we use the following notations:
\begin{equation}
\begin{aligned}
&A=\frac{\sqrt{\mu_a\eta_a}}{2} \\
&B=\frac{\sqrt{\mu_b\eta_b}}{2}
\end{aligned}
\end{equation}
Then, the detection intensities can be simplified and shown in Tab. \ref{tab:detint}. Here, without loss of generality, we let $\theta_c=0$ for simplicity. The physical meaning of $\theta_c$ is the reference phase of $\theta_a$ and $\theta_b$. Therefore, the phase shifting between $\theta_a$ and $\theta_b$ can be compensated by the modulation of $\theta_c$.
\begin{table}[h]
\centering
\caption{\bf Detection intensities on X and Y basis}
\label{tab:detint}
\setlength{\tabcolsep}{12pt}{
\begin{tabular}{ccc}
\hline
$\theta_a$ & $\theta_b$ & $\abs{\psi}^2$ \\
\hline
$0$ & $0$ & $\abs{\psi_1}^2=2A^2$ \\
 &  & $\abs{\psi_2}^2=2B^2$ \\
$\pi$ & $\pi$ & $\abs{\psi_1}^2=2A^2$ \\
 & & $\abs{\psi_2}^2=2B^2$ \\
$0$ & $\pi$ & $\abs{\psi_1}^2=2A^2+2B^2+4AB\cos(\phi_a-\phi_b)$ \\
 & & $\abs{\psi_2}^2=0$ \\
$\pi$ & $0$ & $\abs{\psi_1}^2=0$ \\
 & & $\abs{\psi_2}^2=2A^2+2B^2+4AB\cos(\phi_a-\phi_b)$ \\
 \hline
$\frac{\pi}{2}$ & $\frac{\pi}{2}$ & $\abs{\psi_1}^2=A^2+B^2-2AB\sin(\phi_a-\phi_b)$ \\
 & & $\abs{\psi_2}^2=A^2+B^2+2AB\sin(\phi_a-\phi_b)$ \\
$\frac{3\pi}{2}$ & $\frac{3\pi}{2}$ & $\abs{\psi_1}^2=A^2+B^2+2AB\sin(\phi_a-\phi_b)$ \\
 & & $\abs{\psi_2}^2=A^2+B^2-2AB\sin(\phi_a-\phi_b)$ \\
$\frac{\pi}{2}$ & $\frac{3\pi}{2}$ & $\abs{\psi_1}^2=A^2+B^2+2AB\cos(\phi_a-\phi_b)$ \\
 & & $\abs{\psi_2}^2=A^2+B^2+2AB\cos(\phi_a-\phi_b)$ \\
$\frac{3\pi}{2}$ & $\frac{\pi}{2}$ & $\abs{\psi_1}^2=A^2+B^2+2AB\cos(\phi_a-\phi_b)$ \\
 & & $\abs{\psi_2}^2=A^2+B^2+2AB\cos(\phi_a-\phi_b)$ \\
\hline
\end{tabular}}
\end{table}

The valid response is defined as the coincidence of the clicks of $D_1$ and $D_2$. Therefore, the response probabilities of X basis can be given by
\begin{equation}
\begin{aligned}
\eval{Q_\mu^X}_{\theta_a=\theta_b}=&\sum\limits_{\theta_a=\theta_b\in\{0, \pi\}}\frac{1}{4\pi^2}\int_0^{2\pi}\int_0^{2\pi} p_\mu^{D_1}p_\mu^{D_2} \dd{\phi_a}\dd{\phi_b} \\
=&2(1-(1-Y_0)(1-P_{ap})(2-2A^2-2B^2) \\
&+(1-Y_0)^2(1-P_{ap})^2(1-2A^2)(1-2B^2)) \\
\eval{Q_\mu^X}_{\theta_a\neq\theta_b}=&\sum\limits_{\theta_a\neq\theta_b\in\{0, \pi\}}\frac{1}{4\pi^2}\int_0^{2\pi}\int_0^{2\pi} p_\mu^{D_1}p_\mu^{D_2} \dd{\phi_a}\dd{\phi_b} \\
=&2(1-(1-Y_0)(1-P_{ap})(2-2A^2-2B^2) \\
&+(1-Y_0)^2(1-P_{ap})^2(1-2A^2-2B^2))
\end{aligned}
\end{equation}

Finally, the gains and QBER are given by
\begin{equation}
\begin{aligned}
Q_\mu^X=&\eval{Q_\mu^X}_{\theta_a=\theta_b}+\eval{Q_\mu^X}_{\theta_a\neq\theta_b} \\
E_\mu^XQ_\mu^X=&e_d\eval{Q_\mu^X}_{\theta_a=\theta_b}+(1-e_d)\eval{Q_\mu^X}_{\theta_a\neq\theta_b}
\end{aligned}
\end{equation}
where $e_d$ is the misalignment-error rate.

Similarly, the gains and QBER of Y basis can be given by
\begin{equation}
\begin{aligned}
\eval{Q_\mu^Y}_{\theta_a=\theta_b}=&\sum\limits_{\theta_a=\theta_b\in\{\frac{\pi}{2}, \frac{3\pi}{2}\}}\frac{1}{4\pi^2}\int_0^{2\pi}\int_0^{2\pi} p_\mu^{D_1}p_\mu^{D_2} \dd{\phi_a}\dd{\phi_b} \\
=&2(1-(1-Y_0)(1-P_{ap})(2-2A^2-2B^2) \\
&+(1-Y_0)^2(1-P_{ap})^2((1-A^2-B^2)^2-2A^2B^2))) \\
\eval{Q_\mu^Y}_{\theta_a\neq\theta_b}=&\sum\limits_{\theta_a\neq\theta_b\in\{\frac{\pi}{2}, \frac{3\pi}{2}\}}\frac{1}{4\pi^2}\int_0^{2\pi}\int_0^{2\pi} p_\mu^{D_1}p_\mu^{D_2} \dd{\phi_a}\dd{\phi_b} \\
=&2(1-(1-Y_0)(1-P_{ap})(2-2A^2-2B^2) \\
&+(1-Y_0)^2(1-P_{ap})^2((1-A^2-B^2)^2+2A^2B^2)))
\end{aligned}
\end{equation}
\begin{equation}
\begin{aligned}
Q_\mu^Y=&\eval{Q_\mu^Y}_{\theta_a=\theta_b}+\eval{Q_\mu^Y}_{\theta_a\neq\theta_b} \\
E_\mu^YQ_\mu^Y=&e_d\eval{Q_\mu^Y}_{\theta_a=\theta_b}+(1-e_d)\eval{Q_\mu^Y}_{\theta_a\neq\theta_b}
\end{aligned}
\end{equation}

According to the results shown in Tab. \ref{tab:detint}, when the bit error happens to X basis ($\theta_a\neq\theta_b$ and $\theta_a, \theta_b\in\{0, \pi\}$), one of $\abs{\psi}^2$ is equal to zero. Furthermore, one of $p_\mu$ in Eq. \ref{pd} is close to zero. Therefore, the coincidence probability of error can be very low.

\section{Calculation of secure key rate for MDI-QKD}
\label{app:mdirate}
\setcounter{equation}{0}
\renewcommand{\theequation}{B{\arabic{equation}}}
\setcounter{table}{0}
\renewcommand{\thetable}{B\arabic{table}}
In this section, we use the method of \cite{wang2013three,wang2017measurement} to calculate the secure key rate which treats the statistical fluctuation with Chernoff's bounds. It is enough to show the feasibility of our scheme, although it is not a complete finite-size analysis against the coherent attack.
The secure key rate $R$ can be obtained by
\begin{equation}
R={P_\mu^X}^2(\mu^2e^{-2\mu}Y_{11}^{X,L}(1 - H_2(e_{11}^{Y,U}))-Q^X_{\mu_a \mu_b} f_eH_2((E^X_{\mu_a \mu_b}))\label{mdigllp}
\end{equation}
where $P_\mu^X$ is the probability that Alice and Bob send $\mu$ state with X basis, $Y_{11}^{X,L}$ is the lower bound of the yield of single-photon pairs, $e_{11,p}^{X,U}$ is the upper bound of the phase-flip error rate, $Q^X_{\mu \mu}$ and $E^X_{\mu \mu}$ are the observed gain and QBER that both Alice and Bob send $\mu$ state with X basis, $H_2(x)=-x log_2(x)-(1-x)log_2(1-x)$ is the binary Shannon entropy function and $f_e$ is the error correction efficiency.

The yield and phase-flip error rate can be estimated by observables according to\cite{wang2017measurement}
\begin{equation}
\begin{aligned}
Y_{11}^{X,L}=&\frac {1}{(\mu_a - \omega_a)(\mu_b - \omega_b)(\nu_a - \omega_a)(\mu_b - \omega_b)(\mu_a - \omega_a)} \\
&\times\big[ (\mu_a^2 - \omega_a^2)(\mu_b - \omega_b)(Q^{X,L}_{\nu_a \nu_b} e^{(\nu_a+\nu_b)} + Q^{X,L}_{\omega_a \omega_b}e^{(\omega_a+\omega_b)} \\
&-Q^{X,U}_{\nu_a \omega_b}e^{(\nu_a+\omega_b)} - Q^{X,U}_{\omega_a \nu_b}e^{(\omega_a+\nu_b)} ) \\
&-(\nu_a^2 - \omega_a^2)(\nu_b - \omega_b)(Q^{X,U}_{\mu_a \mu_b} e^{(\mu_a+\mu_b)} + Q^{X,U}_{\omega_a \omega_b}e^{(\omega_a+\omega_b)} \\
&-Q^{X,L}_{\mu_a \omega_b}e^{(\mu_a+\omega_b)} - Q^{X,L}_{\omega_a \mu_b}e^{(\omega_a+\mu_b)} \big]
\end{aligned}
\end{equation}
\begin{equation}
\begin{aligned}
e_{11}^{Y,U}=&\frac {1}{(\nu_a - \omega_a)(\nu_b - \omega_b)Y^{Y,L}_{11}} \\
&\times\big[ e^{(\nu_a + \nu_b)}{EQ}^{Y,U}_{\nu_a\nu_b} + e^{(\omega_a \omega_b)} {EQ}^{Y,U}_{\omega_a\omega_b} \\
&-e^{(\nu_a + \omega_b)}{EQ}^{Y,L}_{\nu_a\omega_b} - e^{(\omega_a + \nu_b)}{EQ}^{Y,L}_{\omega_a\nu_b}  \big]
\end{aligned}
\end{equation}
\begin{equation}
\begin{aligned}
Y_{11}^{Y,L}=&\frac {1}{(\mu_a - \omega_a)(\mu_b - \omega_b)(\nu_a - \omega_a)(\mu_b - \omega_b)(\mu_a - \omega_a)} \\
&\times\big[ (\mu_a^2 - \omega_a^2)(\mu_b - \omega_b)(Q^{Y,L}_{\nu_a \nu_b} e^{(\nu_a+\nu_b)} + Q^{Y,L}_{\omega_a \omega_b}e^{(\omega_a+\omega_b)} \\
&-Q^{Y,U}_{\nu_a \omega_b}e^{(\nu_a+\omega_b)} - Q^{Y,U}_{\omega_a \nu_b}e^{(\omega_a+\nu_b)} ) \\
&-(\nu_a^2 - \omega_a^2)(\nu_b - \omega_b)(Q^{Y,U}_{\mu_a \mu_b} e^{(\mu_a+\mu_b)} + Q^{Y,U}_{\omega_a \omega_b}e^{(\omega_a+\omega_b)} \\
&-Q^{Y,L}_{\mu_a \omega_b}e^{(\mu_a+\omega_b)} - Q^{Y,L}_{\omega_a \mu_b}e^{(\omega_a+\mu_b)} \big].
\end{aligned}
\end{equation}
where $Q^{\beta,\chi}_{\alpha_a \alpha_b}$ and $EQ^{\beta,\chi}_{\alpha_a \alpha_b}$ are the $\chi$ bounds of the observable gain and error rate that Alice sends $\alpha_a$ state and Bob sends $\alpha_b$ state with $\beta$ basis, respectively, $\beta\in\{X, Y\}, \chi\in\{U, L\}, \alpha\in\{\mu, \nu, \omega\}$.

In order to deal with the statistical fluctuation, the observables can be bounded by Chernoff's bounds.
 \begin{equation}
\begin{aligned}
Q^{\beta,U}_{\alpha_a \alpha_b} =&Q^{\beta}_{\alpha_a \alpha_b}(1+\frac{f((\varepsilon/2)^{4}/16)} {\sqrt{N^{\beta}_{\alpha_a \alpha_b}Q^{\beta}_{\alpha_a \alpha_b}}}) \\
Q^{\beta,L}_{\alpha_a \alpha_b} =&Q^{\beta}_{\alpha_a \alpha_b}(1-\frac{f((\varepsilon/2)^{3/2}) }{\sqrt{N^{\beta}_{\alpha_a \alpha_b}Q^{\beta}_{\alpha_a \alpha_b}}}) \\
{EQ}^{\beta,U}_{\alpha_a \alpha_b} =&{EQ}^{\beta}_{\alpha_a \alpha_b}(1+\frac{f((\varepsilon/2)^{4}/16)} {\sqrt{N^{\beta}_{\alpha_a \alpha_b}{EQ}^{\beta}_{\alpha_a \alpha_b}}}) \\
{EQ}^{\beta,L}_{\alpha_a \alpha_b} =&{EQ}^{\beta}_{\alpha_a \alpha_b}(1-\frac{f((\varepsilon/2)^{3/2}) }{\sqrt{N^{\beta}_{\alpha_a \alpha_b}{EQ}^{\beta}_{\alpha_a \alpha_b}}})
\end{aligned}
\end{equation}
where $\varepsilon$ is the failure probability of statistical fluctuation.

The parameters used to calculate the secure key are shown in Table \ref{tab:mdipara}.
\begin{table}[h]
\centering
\caption{\bf Parameters of secure key calculation in our MDI-QKD system}
\label{tab:mdipara}
\setlength{\tabcolsep}{7pt}{
\begin{tabular}{cccccccc}
\hline
$f_e$ & $Y_0$ & $\eta_d$ & $\varepsilon$ & $\mu$ & $\nu$ & $\omega$ \\
$1.16$ & $7.5\times10^{-6}$ & $25\%$ & $10^{-10}$ & 0.284 & 0.057 & 0\\
\hline
\end{tabular}}
\end{table}

\section{Calculation of secure key rate for BB84 QKD}
\label{app:bb84rate}
\setcounter{equation}{0}
\renewcommand{\theequation}{C{\arabic{equation}}}
\setcounter{table}{0}
\renewcommand{\thetable}{C\arabic{table}}

The secure key rate $R$ can be generated by
\begin{equation}
R=\frac{1}{N}(s_0^X+s_1^X(1-H_2(e_{1,p}^X))-\lambda_{EC}-6\log_2\frac{21}{\varepsilon_{sec}}-\log_2\frac{2}{\varepsilon_{cor}})\label{bb84gllp}
\end{equation}
where $s_0$ is the number of vacuum events, $s_1$ is the number of single-photon events, $e_{1,p}$ is the phase error rate, $N$ is the total number of pulses (sent by Alice or Bob), $\beta\in\{X, Y\}$ represents a basis, $H_2(x)=-x log_2(x)-(1-x)log_2(1-x)$ is the binary Shannon entropy function, $\lambda_{EC}=n^\beta f_{e}H_2(E^\beta)$ is the consumption of the information in error-correction, $f_{e}$ is the efficiency factor of the error-correction method used, $\varepsilon_{cor}$ and $\varepsilon_{sec}$ are secure parameters.

All needed parameters can be estimated by analytic formulas\cite{ma2005practical, lim2014concise}. Specifically, the analytic formulas of three-intensity ($\mu, \nu_1, \nu_2$) decoy scheme are given by
\begin{equation}
s_0^\omega=\frac{\tau_0}{\nu_1-\nu_2}\left(\frac{e^{\nu_2}\nu_1 n_{\nu_2}^{\beta,U}}{P_{\nu_2}}-\frac{e^{\nu_1}\nu_2 n_{\nu_1}^{\beta,L}}{P_{\nu_1}}\right)
\end{equation}
\begin{equation}
\begin{aligned}
s_1^\omega=&\frac{\mu\tau_1}{\mu\nu_1-\mu\nu_2-\nu_1^2+\nu_2^2}\bigg[\frac{e^{\nu1}n_{\nu1}^{\beta,L}}{P_{\nu1}}-\frac{e^{\nu2}n_{\nu2}^{\beta,U}}{P_{\nu2}}\\
&-\frac{\nu_1^2-\nu_2^2}{\mu^2}\left(\frac{e^{\mu}n_{\mu}^{\omega,U}}{P_{\mu}}-\frac{s_0^\beta}{\tau_0}\right)\bigg]
\end{aligned}
\end{equation}
\begin{equation}
e_{1,p}^\beta=\frac{v_1^{\overline{\beta}}}{s_1^{\overline{\beta}}}+\gamma\left(\varepsilon_{sec}, \frac{v_1^{\overline{\beta}}}{s_1^{\overline{\beta}}}, s_1^{\overline{\beta}}, s_1^{\beta}\right)
\end{equation}
where
\begin{equation}
\begin{aligned}
\gamma(a,b,c,d)=&\sqrt{\frac{(c+d)(1-b)b\ln2}{cd}}\sqrt{\log_2\left(\frac{c+d}{cd(1-b)b}\frac{21^2}{a^2}\right)}
\end{aligned}
\end{equation}
\begin{equation}
v_1^\beta=\frac{\tau_1}{\nu_1-\nu_2}\left(\frac{e^{\nu1}m_{\nu_1}^{\beta,U}}{P_{\nu_1}}-\frac{e^{\nu2}m_{\nu_2}^{\beta,L}}{P_{\nu_2}}\right)
\end{equation}
$\beta$ and $\overline{\beta}$ are different bases, i.e., $\beta=X$ when $\overline{\beta}=Y$ and vice versa. $n_\alpha^{\beta,\chi}$ and $m_\alpha^{\beta,\chi}$ are the $\chi$ bounds of the number of detections and bit error of basis $\beta$ and intensity $\alpha$, respectively, $\chi\in\{U, L\}, \alpha\in\{\mu, \nu, \omega\}$.

In order to deal with the statistical fluctuation, according to the counterfactual protocol proposed by \cite{lim2014concise}, the counts and errors can be bounded by Hoeffding's inequality.
\begin{equation}
\begin{aligned}
n_\alpha^{\beta,U}=&n_\alpha^\beta+\sqrt{\frac{n^\beta}{2}\ln\frac{21}{\varepsilon_{sec}}} \\
n_\alpha^{\beta,L}=&n_\alpha^\beta-\sqrt{\frac{n^\beta}{2}\ln\frac{21}{\varepsilon_{sec}}} \\
m_\alpha^{\beta,U}=&m_\alpha^\beta+\sqrt{\frac{m^\beta}{2}\ln\frac{21}{\varepsilon_{sec}}} \\
m_\alpha^{\beta,L}=&m_\alpha^\beta-\sqrt{\frac{m^\beta}{2}\ln\frac{21}{\varepsilon_{sec}}}
\end{aligned}
\end{equation}
where $n_\alpha^\omega$ and $m_\alpha^\omega$ are the number of detections and bit error of basis $\omega$ and intensity $\alpha$ observed in experiment.

The parameters used in our secure key calculation are shown in Table \ref{tab:bb84para}.
\begin{table}[h]
\centering
\caption{\bf Parameters of secure key calculation in our BB84 QKD system}
\label{tab:bb84para}
\setlength{\tabcolsep}{5pt}{
\begin{tabular}{cccccccc}
\hline
$f_e$ & $Y_0$ & $\eta_d$ & $\varepsilon_{sec}$ & $\varepsilon_{cor}$ & $\mu$ & $\nu$ & $\omega$ \\
$1.16$ & $7.5\times10^{-6}$ & $25\%$ & $10^{-9}$ & $10^{-15}$ & 0.538 & 0.063 & 0.003\\
\hline
\end{tabular}}
\end{table}
\nocite{*}

\bibliography{sample}

\end{document}